\documentclass[10pt]{article}
 \usepackage[dvips]{graphicx}
 \usepackage{amsmath}
 \usepackage{amsxtra}
\begin{document}
  \bibliographystyle{unsrt}
\title {Energy Dissipation in Quantum Computers}
\bigskip
\author{ A.Granik\thanks{Department of
Physics, University of the
Pacific,Stockton,CA.95211;~E-mail:~agranik@uop.edu}~~and~~G.Chapline\thanks{Lawrence
Livermore National Lab.,
Livermore,CA.94550;E-mail:~Chapline1@LLNL.gov}}
\date{}
\maketitle
\begin{abstract}
A method is described for calculating the heat generated in a
quantum computer due to loss of quantum phase information.
Amazingly enough, this heat generation can take place at zero
temperature. and may explain why it is impossible to extract
energy from vacuum fluctuations. Implications for optical
computers and quantum cosmology are also briefly discussed.
\end{abstract}

During the last decade the literature on  quantum computing grew
so fast and became  so vast, that to review it would require a
separate paper which is not our purpose. We consider the book by
M.Nielsen and I.Chuang \cite{NC} as a well-written and useful
reference text on the topic. Our goal is more modest since we
would like to address one of the problems of a general still
posing a serious obstacle on the path to construction of a
practical quantum computer. The problem is how to prevent
destruction of phase coherence by measurements.\\

In this letter we draw attention to the similarity of such
coherence destroying measurements to logically irreversible
operations in a classical computer. As was first noted by
R.Landauer \cite{RL}, logically irreversible operations {\it
inevitably} result in generation of heat. In particular, Landauer
showed that destruction of $1$ bit of information in a computer
whose environment has temperature $T$ results in a production of
$kTln2$ amount of heat. We will show that destruction of phase
information in a quantum computer also results in a production of
heat.\\

One must be tempted to question whether Landauer's principle
applies to phase information since Shannon's measure of
information , $-\sum_n p_nlnp_n$, depends only on the
probabilities ${p_n}$ for finding the system in various states. On
the other hand, phase information is certainly required to
describe the quantum state of a system. This suggests that the
phase information may reasonably be expected to fall into the
category of "algorithmic information" \cite{WZ} It has been
suggested \cite{CB} that Landauer's principle  does in fact apply
to destruction of algorithmic information.\\

This is certainly true for simple systems in thermal equilibrium
with a hot environment, where ensemble average of algorithmic
information is essential identical with the usual Gibbs-Boltzmann
entropy \cite{CB},\cite{CC}. However, for cold quantum systems the
question seems less clear, since phase information is largely
contained in off-diagonal elements of the density matrix, which
are not directly related to thermodynamic quantities. On the other
hand, energy dissipation in quantum systems can be calculated
directly from the equations of motion for the density matrix.\\

For a closed quantum system there can be no loss of phase
information: indeed, knowledge of phases is exactly equivalent to
knowing how the probabilities for the system to be in various
states change with time. This is illustrated by the Schroedinger
equation written in the following form:
\begin{equation}
\label{1} \hbar\frac{\partial\phi}{\partial
\texttt{}t}+\frac{\hbar^2}{2m}(\nabla\phi)^2+V=\frac{\hbar^2}{2m}\;[\;\frac{1}{2}(\nabla
lnp)^2+\nabla^2(lnp)\;]
\end{equation}
The left-hand side of (\ref{1}) is just the Hamilton-Jacobi
operator for the phase $\phi$ of the wave function $\Psi(x,t)$,
while the r.h.s of (\ref{1}) is a function of the probability
$p(x,t)=|\Psi|^2$ for finding a particle of position $x$.\\

Obviously, a quantum mechanical description  of particle motion
requires a complete knowledge of $\phi(x,t)$. Nevertheless there
are many circumstances where knowledge of $\phi(x,t)$ is
effectively destroyed. For example, measurement of particle's
position  may result in a loss of information concerning
$\phi(x,t)$. Although phase information in the complete system
(particle $\oplus$ measuring apparatus) is presumably conserved
\cite{MS}, there typically will be loss of information concerning
$\phi(x,t)$ when particle's position is measured. We will show in
the following that this loss of phase information is accompanied
by the generation of heat.\\

Our starting point is decoherence functional of Griffiths
\cite{RG}, Omnes{RO}, and Gell-Mann and Hartle \cite{GM}. We
consider a closed quantum system consisting of  "computer"
interacting with an environment. A history for the system is found
by specifying a state $|\alpha(t)>$  for the computer as a
function of time. For each moment of time $t$ one can construct a
projection operator
\begin{equation}
\label{2} P_{\alpha}(t)\equiv
|\alpha(t)><\alpha(t)|\oplus\hat{I}_e
\end{equation}
which projects onto a particular history and leaves the
environment $unchanged$. The projection operators for the same
state at different times are connected by
\begin{equation}
\label{3} P_{\alpha}(t)=e^{Ht/\hbar}P_{\alpha}(0)e^{-Ht/\hbar}
\end{equation}
where $H$ is the total Hamiltonian.\\

The decoherence functional for the system is defined for coarse
grained histories obtained by specifying the states  only at
discrete times. If we denote the corresponding set of projection
operators by $[P_{\alpha}]\equiv
{P_{\alpha_1}^1(t_1),P_{\alpha_2}^2(t_2),...,P_{\alpha_n}^n(t_n)}$
the decoherence functional os then
\begin{equation}
\label{4} D([P_{\alpha'}],[P_{\alpha}])\equiv
Tr[P_{\alpha'_n}^n(t_n)P_{\alpha'_1}^1(t_1)~ \rho_0 ~
P_{\alpha_1}^1(t_1)P_{\alpha_n}^n(t_n)]
\end{equation}
where $\rho_0$ is the initial density matrix. In cases where
$\rho_0$ can be factored into a density matrix for the system and
a density matrix for the environment, the decoherence  functional
can be expressed in terms of the influence functional of Feynman
and Vernon \cite{RF}. This allows one to explicitly calculate the
effects of phase decoherence when the environment consists of
harmonic oscillators, since the influence functional can be
explicitly calculated in this case \cite{RF}.\\

Although the coarse graining of time histories is quite
appropriate for computers, where the results of calculations are
normally displayed at discrete times, it would be convenient for
illustrative purposes to assume that our "computer" also consists
of an array of harmonic oscillators whose amplitudes are monitored
$ continuously$ in time and specified by paths $x_i(t)$. Although
this example differs conceptually from the conceptions of quantum
computers discussed , for example, in some pioneering theoretical
models of such computers \cite{PB},\cite{RPF},\cite{DD},\cite{AP}
it is not without interest in itself. In particular, one might
think of our array of harmonic oscillators as the eigenmodes of an
optical resonator.\\

In optical realizations of associative memory \cite{DA}, \cite{BH}
information about this modes is typically stored holographically,
and depending on the similarity of an input signal to one of the
stored modes the optical system will resonate in one of the
eigenmodes. In a quantum mechanical version of such an optical
system the state of the system would be described by a wave
function $\Psi(x,t)$ where $x$ denotes the amplitude of the
electric field corresponding to a particular eigenmode. Actually,
optical versions of associative memories will typically include
non-linear elements providing gain for the oscillators. Therefore
the problem of describing (negative) dissipation arises out of
necessity. Fortunately, the influence functional formalism is
capable of describing the effects of feedback and gain. However,
our main purpose here is not to describe the well-understood
amplification of a coherent signal in a medium with gain, but
instead to study energy dissipation associated with loss of phase
information for $\Psi(x,t)$.\\

Assuming that the initial density matrix $\rho_0$ for a harmonic
oscillator plus environment ca be factorized, we obtain for the
decoherence functional for a time interval $t=0$ to $t=t_f$ the
following expression:
\begin{equation}
\label{5}
D(x(t),y(t))=\delta[x(t_f)-y(t_f)]~e^{i[S(x(t))-S(y(t))]}~e^{iW(x,y)}\rho(y(0),x(0))
\end{equation}
where $\rho(x(0),y(0))$ is the initial density matrix for the
harmonic oscillator. The effect of the environment on the harmonic
oscillator is summarized by the influence functional $e^{iW}$. If
the harmonic oscillator is linearly coupled to the environmental
harmonic oscillators then $W$ has the form \cite{RPF}
\begin{equation}
\label{6}
W=i\int_0^{t_f}\int_0^{t_f}ds[x(t)-y(t)][\alpha(t-s)x(s)-\alpha^*(t-s)y(s)]
\end{equation}

We will use in (\ref{6}) an expression for  $\alpha$ chosen on the basis of\\

a) a continuum of oscillators representing the thermal bath\

 and\

b) the representation of oscillator's density according to Zwanzig
\cite{RZ}. \\

Given these assumptions, the real and imaginary parts of $\alpha$
have the form
\begin{equation}
\label{7} \alpha_R=\frac{\eta}{\pi}\int_0^\Omega\omega
coth\frac{\hbar\omega}{2k_BT}cos\omega(t-s)d\omega
\end{equation}
\begin{equation}
\label{8} \alpha_I=\frac{\eta}{2\pi}\int_0^\Omega\omega
sin\omega(t-s)d\omega=\frac{\eta}{\pi}\frac{\Omega}{t-s}[
cos\Omega(t-s)-\frac{sin\Omega(t-s)}{\Omega(t-s)}]
\end{equation}
where $\Omega$ is a high frequency cut-off and $\eta$ is a damping
coefficient.\\

The integral in (\ref{7}) cannot be evaluated in a closed form.
However, for the case which we are considering here, i.e.,
$T\rightarrow 0$, $\alpha_R$ is easily found. This is achieved by
replacing $coth(\omega/2k_BT)$ ($k_B$ is the Boltzmann constant)
with $1$ for all the frequencies $0\le\omega\le\Omega$. We include
$\omega=0$, because at this point the integrand is $0$, which is
easily seen from the following:\\

 $a) \;\;\;\mathop {\mathop {Lim}\limits_{\omega  \rightarrow 0} }\limits_{T \to
0}\displaystyle \frac{\hbar\omega}{k_BT}=0\;\;\;\;\;\;
\rightarrow\;\;\;\;\; \mathop {\mathop {Lim}\limits_{\omega
\rightarrow 0} }\limits_{T \to 0}\omega coth \frac{\hbar\omega}{
2k_BT}=\frac{\omega}{\hbar\omega/2k_BT}=\frac{2k_BT}{\hbar}=0$\

$b)\;\;\;\mathop {\mathop {Lim}\limits_{\hbar\omega  \rightarrow
0} }\limits_{T \to 0}\displaystyle\frac{\hbar\omega}{k_BT}=C\neq 0
\rightarrow \;\;\;\;\mathop {\mathop {Lim}\limits_{\omega
\rightarrow 0} }\limits_{T \to 0}\omega coth \frac{\hbar\omega}{
2k_BT}=\omega coth\frac{C}{2}=0$\

$c)\;\;\;\mathop {\mathop {Lim}\limits_{\omega  \rightarrow 0}
}\limits_{T \to 0}\displaystyle
\frac{\hbar\omega}{k_BT}\rightarrow\infty\;\;\;\;\;\rightarrow\;\;\;\;\mathop
{\mathop {Lim}\limits_{\omega  \rightarrow 0} }\limits_{T \to
0}\displaystyle \omega coth\frac{\hbar\omega}{2k_BT}=\omega=0$\\
\

As a result, we obtain from (\ref{7})
\begin{equation}
\label{9} \alpha_R(T=0)=\frac{\eta}{\pi} \Omega\frac
{sin\Omega(t-s)}{t-s}-\frac{\eta}{2\pi}\Biggl(\frac{sin\frac{\Omega(t-s)}{2}}{\frac{t-s}{2}}\Biggr)^2
\end{equation}
The imaginary part of $W$ (Eq.\ref{6}) is then
\begin{eqnarray}
\label{10} W_I=\frac{\eta}{\pi}\Biggl\{\int_0^t\int_0^\tau
[x(t)-y(t)][x(s)-y(s)]\frac{\Omega sin\Omega(t-s)}{t-s}dtds\nonumber\\
-\frac{1}{2}\int_0^t\int_0^\tau
[x(t)-y(t)][x(s)-y(s)]\Biggl(\frac{sin\frac{\Omega(t-s)}{2}}{\frac{t-s}{2}}\Biggr)^2dsdt\Biggr\}
\end{eqnarray}
If $\Omega\to \infty$ then $\displaystyle\frac{1}{\pi}
sin\frac{\Omega(t-s)}{(t-s)}=\delta(t-s)$, and Eq.(\ref{10})
yields
\begin{equation}
\label{11}
W_I=[\eta\Omega-\frac{\pi\eta}{2}\delta(0)]\int_0^t[x(t)-y(t)]^2dt
\end{equation}
Taking into account the fact that
$\displaystyle\lim_{\Omega\to\infty}\frac{\Omega}{\pi}=\delta(0)$
\footnote{the detailed discussion of the validity of this limit is
provided in \cite{AC}}, we obtain from (\ref{11})
\begin{equation}
\label{12}
\lim_{\Omega\to\infty}\frac{W_I}{\hbar}=\frac{\eta\Omega}{2\hbar}\int_0^t[x(t)-y(t)]^2dt
\end{equation}
This expression allows us to estimate the decoherence time
(corresponding to $\displaystyle \frac{W_I}{\hbar}\approx 1$) at
zero temperature $T=0$:
\begin{equation}
\label{13}
t_d\approx\frac{1}{\eta\Omega}\frac{\hbar}{|x(0)-y(0)|^2}
\end{equation}
If we compare $t_d$ with the analogous expression for
$k_BT>>\hbar\Omega$ \cite{AC}, namely
$$t_d\approx\frac{\hbar}{\eta k_BT}\frac{\hbar}{|x(0)-y(0)|^2}$$ we
can see that $\hbar\Omega$ now plays the role of the "temperature"
\footnote{probably it can be called the decoherence
"temperature"}.\\

In order to calculate the energy transfer from our oscillator
computer to its environment due to loss of phase information, we
can treat the term in Eq.(\ref{5}) arising from $\alpha_R$ as a
time dependent perturbation. If we neglect the effect of
$\alpha_I$, the equation of motion for the oscillator density
matrix in an unperturbed eigenstate basis is
\begin{equation}
\label{14}
\frac{d\rho_{nm}}{dt}=-\frac{i}{\hbar}\omega_{nm}\rho_{nm}-
\frac{\eta\Omega}{\pi\hbar}<n|[\;x,[x,\rho]\;]|m>
\end{equation}\

To first order in the interaction we find that the change in
probability for finding the computer in its $n-th$ energy
eigenstate is
\begin{equation}
\label{15}
\Delta\rho_{nn}\simeq-\frac{\eta\Omega}{\pi\hbar}\int_0^{\infty}dt
\int\int dxdy\;\rho_0(x,y)\psi_n^*(x)(x-y)^2\psi_n(y)
\end{equation}
where $\rho_0(x,y)$ is the initial density matrix. From
Eq.(\ref{15}) it is straightforward to calculate the expected
energy loss $$\Delta E=\displaystyle\sum_n\Delta\rho_{nn}E_n$$
Clearly, the same sort of procedure will apply to $any$ quantum
computer, with an appropriate operator replacing $x$ in
Eq.(\ref{14}). Thus we have found a general method for calculating
the energy loss associated with loss of phase information.\\

However, we have to inject a word of caution here. It should be
noted that we have $not$ found a general formula for the amount of
phase information that has been dissipated. Therefore, it is not
possible to compare our result directly with Landauer's principle.
It is interesting that our formula for energy loss, Eq.(\ref{15}),
does not involve temperature at all, and hence one might be
tempted to conclude that our result does not agree with Landauer's
principle. However , injecting any amount of heat into a cold
environment will raise its temperature, and so the question of
whether our result agrees with Landauer's principle requires a
more careful consideration.

The fact that energy dissipation can occur at $ zero$ temperature
is nonetheless a unique result, and may have some bearing on the
question whether it is possible to extract energy from quantum
fluctuations of vacuum. Bennett \cite {CB} has pointed out that
Landauer's principle lies at the heart of the Maxwell Demon
problem: i.e., why it is impossible for information gathering
system to extract energy from thermal fluctuations. One might
expect that some kind of quantum analog of Landauer's principle
would explain why it is impossible to extract energy from quantum
fluctuations. Since a detailed description of quantum fluctuations
would almost certainly involve phase information, the question of
energy cost of erasing this information would very likely to play
a decisive role. Furthermore, since quantum fluctuations can take
place at zero temperature one would expect that the energy cost of
erasing phase information should also $not$ vanish at zero
temperature, as is indicated by the above analysis.\\

The inevitable loss of phase information implied by
Eqs.(\ref{5})-(\ref{13}) does have certain implications for the
possibility of using quantum effects in optical computing. Optical
computers (e.g.,\cite{JC}) normally use coherent states for the
electromagnetic field. Roughly speaking, such a use of coherent
states is equivalent to the requirement that every photon perform
the same calculation. Apparently, greater computational power
would become available if more general states for the photons,
e.g., squeezed states, could be used. Unfortunately,
Eqs(\ref{5})-(\ref{13}) suggest that this is not possible in
practice. As noted above, optical computers typically incorporate
non-linear devices to provide gain. The usual formulae \cite{AC}
describing decoherence due to thermal fluctuations become
meaningless in the presence of such devices. However, our quantum
decoherence formulae are presumably valid and imply that
decoherence will occur in a time on the order of the given in
Eq.(\ref{13}). where $\Omega$ is on the order of the photon
frequency. This will be a very short time for high gain devices
employing visible light. For example, in an optical computer using
$N$ photons and a non-linear amplifier with a gain $1\; cm^{-1}$,
the decoherence time $\displaystyle t_d\approx\frac{10^{-10}}{N}\;
sec$.\\

A formally similar problem concerns the question of how classical
space-time emerges in quantum cosmology. As in the photon case,
the classical field description for the metric of space-time
represents a special state for the quantized gravitational field
corresponding to loss of phase information \cite{JH}. Evidently,
the initial state and environment for the gravitational field are
such that the off-diagonal elements of the decoherence functional
become small. In the twistor formulation of quantum gravity
\cite{GC} the Lagrangian for the gravitational field has a form
very similar to the usual electromagnetic field-atomic
interaction, with certain twistor fields playing the role of
atoms. Thus the behavior of quantum optical devices may provide a
model for quantum cosmology. For example, the main theme of this
short paper, the energy dissipation associated with loss of
quantum information, should have an analog in quantum  cosmology,
In fact, it is very tempting to identify the heat generated as a
result of the loss of phase information with $3^0$ cosmic
blackbody radiation.

\end{document}